\documentclass[aps,prl,twocolumn,superscriptaddress,showpacs,floatfix]{revtex4}

\usepackage{graphicx}
\usepackage{dcolumn}
\usepackage{bm}
\usepackage{hyperref}
\usepackage{amssymb,amsmath}
\IfFileExists{txfonts.sty}{\usepackage{txfonts}}{}

\newcommand{\Dspm}{\ensuremath{D_s^\pm}}
\newcommand{\Dsmp}{\ensuremath{D_s^\mp}}
\newcommand{\D}{\ensuremath{D}}
\newcommand{\Dbar}{\ensuremath{\overline{D}}}
\newcommand{\Ds}{\ensuremath{D_{s}}}
\newcommand{\Dsp}{\ensuremath{D_{s}^+}}
\newcommand{\Dsm}{\ensuremath{D_{s}^-}}

\newcommand{\mrec}{\ensuremath{M_\mathrm{rec}}}

\newcommand{\Dz}{\ensuremath{D^0}}

\newcommand{\Dp}{\ensuremath{D^+}}

\newcommand{\Kp}{\ensuremath{K^+}}
\newcommand{\Km}{\ensuremath{K^-}}
\newcommand{\KS}{\ensuremath{K_{S}^0}\,}
\newcommand{\pip}{\ensuremath{\pi^+}}
\newcommand{\pim}{\ensuremath{\pi^-}}
\newcommand{\piz}{\ensuremath{{\pi^0}}}
\newcommand{\Jpsi}{\ensuremath{J/\psi}}
\newcommand{\invpb}{pb$^{-1}$}
\newcommand{\Kstarzb}{\ensuremath{\overline{K}^{*0}}}
\newcommand{\Br}{\ensuremath{\mathcal{B}}}
\newcommand{\cleoc}{\hbox{CLEO-c}}

\newcommand{\BFKSK}{\ensuremath{1.49 \pm 0.07 \pm 0.05}}
\newcommand{\BFKKpi}{\ensuremath{5.50 \pm 0.23 \pm 0.16}}
\newcommand{\BFKKpipiz}{\ensuremath{5.65 \pm 0.29 \pm 0.40}}
\newcommand{\BFKSKpipi}{\ensuremath{1.64 \pm 0.10 \pm 0.07}}
\newcommand{\BFpipipi}{\ensuremath{1.11 \pm 0.07 \pm 0.04}}
\newcommand{\BFpieta}{\ensuremath{1.58 \pm 0.11 \pm 0.18}}
\newcommand{\BFpietaprime}{\ensuremath{3.77 \pm 0.25 \pm 0.30}}
\newcommand{\BFKpipi}{\ensuremath{0.69 \pm 0.05 \pm 0.03}}

\newcommand{\BRKSK}{\ensuremath{0.270 \pm 0.009 \pm 0.008}}
\newcommand{\BRKKpipiz}{\ensuremath{1.03 \pm 0.05 \pm 0.08}}
\newcommand{\BRKSKpipi}{\ensuremath{0.298 \pm 0.014 \pm 0.011}}
\newcommand{\BRpipipi}{\ensuremath{0.202 \pm 0.011 \pm 0.009}}
\newcommand{\BRpieta}{\ensuremath{0.288 \pm 0.018 \pm 0.033}}
\newcommand{\BRpietaprime}{\ensuremath{0.69\pm 0.04 \pm 0.06}}
\newcommand{\BRKpipi}{\ensuremath{0.125 \pm 0.009 \pm 0.005}}

\newcommand{\AKSK}{\ensuremath{+4.9 \pm 2.1 \pm 0.9}}
\newcommand{\AKKpi}{\ensuremath{+0.3 \pm 1.1 \pm 0.8}}
\newcommand{\AKKpipiz}{\ensuremath{-5.9 \pm 4.2 \pm 1.2}}
\newcommand{\AKSKpipi}{\ensuremath{-0.7 \pm 3.6 \pm 1.1}}
\newcommand{\Apipipi}{\ensuremath{+2.0 \pm 4.6 \pm 0.7}}
\newcommand{\Apieta}{\ensuremath{-8.2 \pm 5.2 \pm 0.8}}
\newcommand{\Apietaprime}{\ensuremath{-5.5 \pm 3.7 \pm 1.2}}
\newcommand{\AKpipi}{\ensuremath{+11.2 \pm 7.0 \pm 0.9}}

\newcommand{\BFphipiv}{\ensuremath{1.69 \pm 0.08 \pm 0.06}}
\newcommand{\BFphipix}{\ensuremath{1.99 \pm 0.10 \pm 0.05}}
\newcommand{\BFphipixv}{\ensuremath{2.14 \pm 0.10 \pm 0.05}}
\newcommand{\BFphipixx}{\ensuremath{2.24 \pm 0.11 \pm 0.06}}

\newcommand{\NDsDs}{\ensuremath{(2.93 \pm 0.14 \pm 0.06) \times 10^5}}
\newcommand{\sigDsDs}{\ensuremath{0.983 \pm 0.046 \pm 0.021 \pm 0.010}}

\begin{document}
\preprint{CLNS 07/2016}       
\preprint{CLEO 07-20}         

\title{\boldmath Absolute Measurement of Hadronic Branching Fractions of the \Dsp\ Meson}

\author{J.~P.~Alexander}
\author{K.~Berkelman}
\author{D.~G.~Cassel}
\author{J.~E.~Duboscq}
\author{R.~Ehrlich}
\author{L.~Fields}
\author{L.~Gibbons}
\author{R.~Gray}
\author{S.~W.~Gray}
\author{D.~L.~Hartill}
\author{B.~K.~Heltsley}
\author{D.~Hertz}
\author{C.~D.~Jones}
\author{J.~Kandaswamy}
\author{D.~L.~Kreinick}
\author{V.~E.~Kuznetsov}
\author{H.~Mahlke-Kr\"uger}
\author{D.~Mohapatra}
\author{P.~U.~E.~Onyisi}
\author{J.~R.~Patterson}
\author{D.~Peterson}
\author{D.~Riley}
\author{A.~Ryd}
\author{A.~J.~Sadoff}
\author{X.~Shi}
\author{S.~Stroiney}
\author{W.~M.~Sun}
\author{T.~Wilksen}
\affiliation{Cornell University, Ithaca, New York 14853, USA}
% ...
\collaboration{CLEO Collaboration} 
\noaffiliation


\date{December 30, 2007}

\begin{abstract} 
The branching fractions of \Dspm\ meson decays serve to normalize many measurements of processes involving charm quarks.  Using 298~\invpb\ of $e^+ e^-$ collisions recorded at a center of mass energy of 4.17~GeV, we determine absolute branching fractions for eight \Dspm\ decays with a double tag technique.  In particular we determine the branching fraction $\Br(\Dsp \to \Km\Kp\pip) = (\BFKKpi)\%$, where the uncertainties are statistical and systematic respectively.  We also provide partial branching fractions for kinematic subsets of the $\Km\Kp\pip$ decay mode. 
\end{abstract}

\maketitle

Uncertainties in the decay probabilities (branching fractions) of the \Dsp\ meson to various detectable final states significantly impact the precision of a diverse array of measurements, including tests of the Standard Model prediction of the coupling of the $Z^0$ boson to charm quarks, measurements of $B$ meson properties such as $B^0_s$ mixing parameters, tests of light quark SU(3) symmetry in the $D$ system, and tests of lattice gauge theory in leptonic \Dsp\ decays.
Any rate measurement where a \Dsp\ meson is an intermediate step in a decay chain demands that the relevant normalizing branching fractions be known precisely to reduce systematic uncertainties.  Most \Dsp\ branching fractions are presently obtained by combining measurements of ratios with a single absolute branching fraction of one decay mode, thus introducing strong correlations and compounding uncertainties.  In this Letter we present the first simultaneous high-statistics determination of multiple \Dsp\ absolute branching fractions, using a technique with significantly different systematic uncertainties from previous branching fraction measurements, which results in precision better than current world averages by a factor of two.
The eight decays considered in this analysis are $\Dsp \to \KS\Kp$, $\Km\Kp\pip$, $\Km\Kp\pip\piz$, $\KS \Km\pip\pip$, $\pip\pip\pim$, $\pip\eta$, $\pip\eta'$, and $\Kp\pip\pim$.  Except where noted, mention of a decay implies the charge conjugate process as well.

The most precise measurements of absolute \Ds\ branching fractions are currently obtained using partial reconstruction techniques to obtain the total number of \Ds\ mesons produced, either from $B \to D^{(*)} D_{s(J)}^{(*)+}$ decays \cite{Aubert:2005xu,Aubert:2006nm} or from $e^+ e^- \to \Ds^{*\pm} D_{s1}^{\mp}(2536)$ events \cite{Abe:2007jz}.  References \cite{Aubert:2005xu} and \cite{Aubert:2006nm} quote results for the resonant decay $\Dsp \to \phi \pip$, while Ref.~\cite{Abe:2007jz} measures $\Br(\Dsp \to \Km\Kp\pip)$.

Here we employ a technique extensively used by \cleoc, pioneered by the MARK~III collaboration for measuring \Dz\ and \Dp\ branching fractions \cite{Baltrusaitis:1985iw,Adler:1987as} and limiting \Ds\ branching fractions \cite{Adler:1989st}, which exploits a feature of near-threshold production of charmed mesons.  Below the $\Ds D K$ threshold of 4.33~GeV, production of a \Dspm\ meson in a charm- and strangeness-conserving process requires the existence of a \Dsmp\ meson elsewhere in the event (possibly with additional photons or pions).  Events where at least one \Ds\ candidate is reconstructed (``single tag'' or ST events) thus provide a sample with a known number of \Ds\ events.  Absolute branching fractions can then be obtained by finding events with two reconstructed \Ds\ candidates (``double tag'' or DT events).  In this analysis, yields for charge-conjugate ST modes are considered separately, but charge-conjugate branching fractions are assumed to be equal, ignoring the possibility of direct $CP$ violation.  There are 16 ST yields, corresponding to two charges for each considered \Ds\ decay, and 64 DT yields, one for each pairing of a \Dsm\ and a \Dsp\ decay.

This analysis uses $(298 \pm 3)$~\invpb\ of data taken at a center of mass energy of 4.17~GeV.  At this energy the dominant \Ds\ production mechanism is the process $e^+ e^- \to \Ds^{*\pm}\Dsmp$ with a cross-section of $\sim 1$ nb \cite{CroninHennessy:2008yi}; the $\Ds^*$ then decays to either $\gamma \Ds$ or $\piz \Ds$ in a $\sim 16:1$ ratio \cite{Yao:2006px}.  The very small rate of $e^+ e^- \to \Dsp\Dsm$ is not used for this analysis.  The transition photon or $\piz$ is not reconstructed.

To illustrate the method, consider two ST modes, $\Dsp \to i$ and $\Dsm \to \bar \jmath$, and one DT mode $\Dsp \to i$, $\Dsm \to \bar \jmath$.  For a given number of produced \Ds\ pairs $N_{\Ds^*\Ds}$, branching fractions $\Br_i$ and $\Br_j$, and efficiencies for the ST ($\epsilon_i$ and $\epsilon_{\bar \jmath}$) and DT ($\epsilon_{i\bar \jmath}$) events, we expect to observe the yields
\begin{align*}
y_i  & = N_{\Ds^*\Ds} \Br_i \epsilon_i,\\
y_{\bar \jmath}  & = N_{\Ds^*\Ds} \Br_j \epsilon_{\bar \jmath},\\
y_{i\bar \jmath} & = N_{\Ds^*\Ds} \Br_i \Br_j \epsilon_{i\bar \jmath},
\end{align*}
where $y_i$ and $y_{\bar \jmath}$ are the ST yields and $y_{i\bar \jmath}$ is the DT yield.
Using $\epsilon_i$, $\epsilon_{\bar \jmath}$, and $\epsilon_{i\bar \jmath}$ from Monte Carlo simulations, we can solve for the branching fractions and $N_{\Ds^*\Ds}$:
\begin{align*}
\Br_i &= \frac{y_{i\bar \jmath}}{y_{\bar \jmath}} \frac{\epsilon_{\bar \jmath}}{\epsilon_{i\bar \jmath}},\\
\Br_j &= \frac{y_{i\bar \jmath}}{y_{i}} \frac{\epsilon_{i}}{\epsilon_{i\bar \jmath}}, \\
N_{\Ds^*\Ds} &= \frac{y_i y_{\bar \jmath}}{y_{i \bar \jmath}} \frac{\epsilon_{i\bar \jmath}}{\epsilon_{i} \epsilon_{\bar \jmath}}.
\end{align*}
In practice, to maximize the statistical power of the analysis, the parameters $N_{\Ds^*\Ds}$ and $\Br_i$ are simultaneously optimized using a maximum likelihood fit to the observed yields, where the ST yields use Gaussian likelihood functions and the considerably smaller DT yields use Poisson likelihood functions.  The statistical properties of the fit were checked with pseudoexperiments.

The \cleoc\  detector is a modification of the CLEO~III detector \cite{cleoiidetector,cleoiiidr,Artuso:2005dc,cleocyb}.  The silicon strip vertex detector was replaced by a six layer vertex drift chamber.  The charged particle tracking system, consisting of the vertex chamber and a \hbox{47-layer} central drift chamber, operates in an axial 1~T magnetic field, and provides a momentum resolution $\sigma_p/p \sim 0.6\%$ at $p = 1\textnormal{ GeV}/c$ for tracks traversing every layer.  Photons are detected in an electromagnetic calorimeter consisting of 7784 CsI(Tl) crystals, which achieves an energy resolution $\sigma_E/E \sim 5\%$ at 100~MeV.  Two particle identification (PID) systems are used to distinguish charged kaons and pions: the central drift chamber, which provides specific ionization measurements ($dE/dx$), and, surrounding this chamber, a cylindrical Ring Imaging Cherenkov (RICH) detector.  The combined PID system achieves $\pi^\pm$ and $K^\pm$ identification efficiency exceeding $85\%$ with fake rates less than $5\%$ over the kinematic range of interest \cite{Dobbs2007}.  The detector response is modeled with a detailed \textsc{geant}-based \cite{geant} Monte Carlo simulation, with initial particle trajectories generated by \textsc{evtgen} \cite{evtgen} and final state radiation produced by \textsc{photos} \cite{photos}.  The initial state radiation spectrum is modeled using cross-sections for $\Ds^*\Ds$ production at lower energies determined during a \cleoc\  scan of this region \cite{CroninHennessy:2008yi}.

Charged tracks are required to be well-reconstructed and, except for \KS\ daughters, to be consistent with originating at the interaction point.  The initial selection requires track momentum $> 50$~MeV$/c$.  Candidate $K^\pm$ and $\pi^\pm$ tracks are chosen using $dE/dx$ and RICH information, using the same criteria as the \cleoc\  \Dz/\Dp\ absolute branching fraction analysis~\cite{Dobbs2007}.  Charged kaons must have momentum above $125\textnormal{ MeV}/c$.   Neutral kaon candidates are reconstructed in the $\KS \to \pip\pim$ decay.  The two pions have no PID requirements, and a vertex fit is done to allow for the \KS\ flight distance.  The pion pair is required to have an invariant mass within $6.3\textnormal{ MeV}/c^2$ of the nominal \KS\ mass.  We form \piz\ and $\eta$ candidates using pairs of isolated electromagnetic showers, keeping combinations within three standard deviations of the nominal masses; for further use a kinematic fit constrains the candidates to the nominal mass.  Candidate $\eta'$ mesons are reconstructed by combining $\eta$ candidates with $\pip\pim$ pairs; the pions are subject to the standard pion PID requirements, and the reconstructed $\eta'$ mass must be within $10\textnormal{ MeV}/c^2$ of the nominal value.

We use several samples of simulated events to obtain efficiencies, study background shapes, and cross-check the analysis.  A ``generic'' decay models a physical \Ds\ decaying into any of its final states; the branching fractions and intermediate resonant components used for various final states are motivated by Particle Data Group (PDG) averages \cite{Yao:2006px}.  A ``signal'' decay is one in which the simulated \Ds\ always decays to a final state of interest, with the same ratio of resonant components as in generic decays.  We obtain efficiencies from samples with either one signal and one generic decay (ST modes) or two signal decays (DT modes).  Backgrounds are investigated using a combined sample of generic \Dz, \Dp, and \Ds\ decays with appropriate production mechanisms and rates at 4.17 GeV, and samples of $e^+ e^- \to \tau^+ \tau^-$, $\gamma \psi(2S)$, and light quarks.

We identify \Ds\ candidates using their momenta and invariant masses.  A candidate may either be the daughter of a $\Ds^*$ (an ``indirect'' \Ds) or be produced in the inital $e^+ e^- \to \Ds^*\Ds$ process (a ``direct'' \Ds).  Direct \Ds\ candidates have fixed momentum in the center of mass frame because they are produced in a two-body process; indirect \Ds\ candidates have a momentum distribution smeared around this value due to the extra boost of the $\Ds^* \to (\gamma,\piz)\Ds$ decay.  We define the recoil mass variable \mrec\ through
\[ \mrec^2 c^4 \equiv \left(E_0 - \sqrt{\mathbf{p}_{\Ds}^2 c^2 + M_{\Ds}^2 c^4}\right)^2 - (\mathbf{p}_0 - \mathbf{p}_{\Ds})^2 c^2, \]
where $(E_0, \mathbf{p}_0)$ is the $e^+ e^-$ center of mass four-vector, $\mathbf{p}_{\Ds}$ is the measured \Ds\ momentum, and $M_{\Ds}$ is the nominal \Ds\ mass.  For direct \Ds\ candidates, \mrec\ peaks at the $\Ds^*$ mass of 2.112~GeV$/c^2$; for indirect \Ds\ candidates, \mrec\ spreads roughly $\pm 60\textnormal{ MeV}/c^2$ around this peak.  For DT and most ST candidates, we require $\mrec > 2.051\textnormal{ GeV}/c^2$; this accepts all kinematically allowed events.  For three ST modes ($\Km\Kp\pip\piz$, $\pip\pip\pim$, $\Kp\pip\pim$) tighter mode-dependent selections of $\mrec > (2.099, 2.101, 2.099)\textnormal{ GeV}/c^2$, which are roughly 80\% efficient for signal, are applied to improve the signal to background ratio.  The \mrec\ requirement eliminates contributions from $e^+ e^- \to \Dsp\Dsm$ events as those occur in a narrow peak at $\mrec = M_{\Ds}$.

The \Ds\ candidates are subject to mode-dependent vetoes to reduce structure in the background invariant mass spectrum, mostly arising from copiously produced $D^* D^*$ events.  In all modes except $\KS \Kp$ and $\Km\Kp\pip$, all neutral and charged pions, including \KS\ daughters, must have momentum above $100\textnormal{ MeV}/c$ to eliminate the soft pions from $D^*$ decays.  Reflections are reduced by vetoing candidates where certain daughter combinations are consistent with the \Dz\ or \Dp\ masses (for example, the $\Km\Kp$ pair in a $\Km\Kp\pip$ candidate must not be consistent with a Cabibbo-suppressed \Dz\ decay).  To remove contamination from \KS\ decays in the modes $\pip\pip\pim$ and $\Kp\pip\pim$, no $\pip\pim$ combination may have a mass between 475 and 520~MeV$/c^2$.

\begin{figure}
\includegraphics*[width=\linewidth]{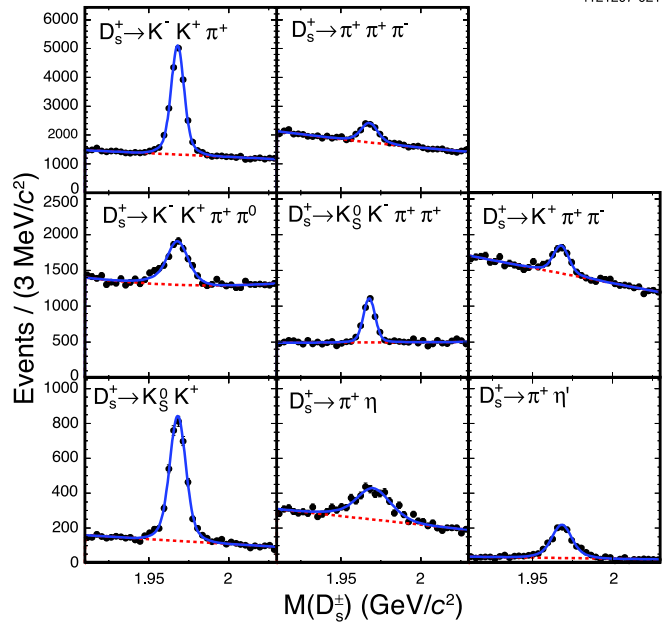}
\caption{\label{fig:stall}(Color online) Invariant masses of the \Dspm\ candidates in data in ST modes.  Charge-conjugate modes are combined.  The fits for yields are shown as the dashed red lines (background component) and solid blue lines (signal plus background).  The total ST yield is $(30.9 \pm 0.3)\times 10^3$ events.}
\end{figure}

For ST yield extraction, every event is allowed to contribute a maximum of one \Ds\ candidate per mode and charge.  If there are multiple candidates, the one with \mrec\ closest to $M_{\Ds^*}$ is chosen.  An unbinned maximum likelihood fit is then performed on the invariant mass spectrum of the candidates in each mode.  The expected signal distribution is obtained from Monte Carlo simulations; in fits to data the \Ds\ mass is allowed to float.
Backgrounds are modeled with linear functions for all modes except $\Km\Kp\pip\piz$ and $\pip\pip\pim$, where quadratic functions are used.  The same background shape is used for both charges in a given mode.  The reconstructed candidate masses $M(\Ds)$ and ST yield fits are shown in Fig.~\ref{fig:stall}.  Efficiencies for ST modes range from 5.3\% to 51\%.

\begin{figure}
\includegraphics*[width=\linewidth]{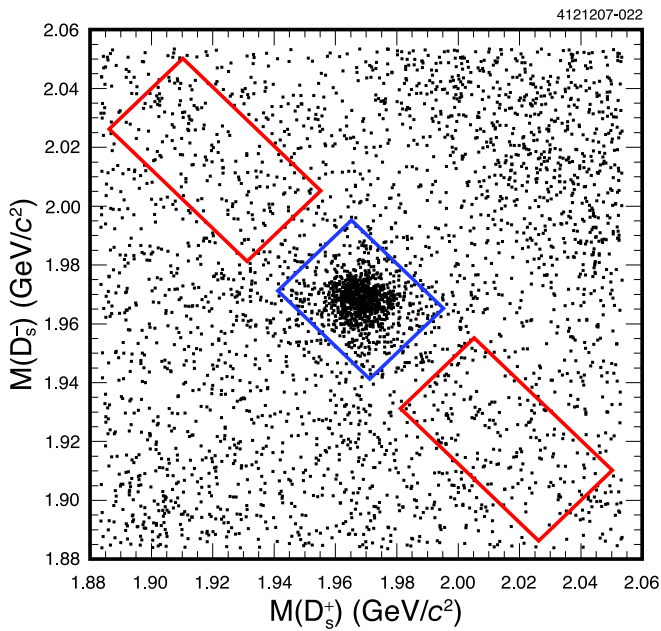}
\caption{\label{fig:dtall}(Color online) Masses of the \Dsm\ and \Dsp\ candidates for all 64 DT modes in data.  The rectangles show the signal region (center) and two sideband regions (diagonally offset).  There are 1089 events in the signal region and 339 events in the combined sideband regions.}
\end{figure}

Double tag yields are extracted by defining a signal region in the two-dimensional plane of the two \Ds\ candidate masses, $M(\Dsp)$ vs.~$M(\Dsm)$.  
Every event is allowed to contribute at most one DT candidate per possible final state; amongst multiple candidates, the combination with average mass $\widehat M = (M(\Dsp) + M(\Dsm))/2$ closest to $M_{\Ds}$ is chosen.  The distribution of $M(\Dsm)$ versus $M(\Dsp)$ for all DT candidates, along with the signal and sideband regions, is shown in Fig.~\ref{fig:dtall}.  Combinatoric background tends to have structure in $\widehat M$ but be flat in the mass difference $\Delta M = M(\Dsp) - M(\Dsm)$; in particular simulations verify that the multiple candidate selection does not cause backgrounds to peak in $\Delta M$.  Both signal and sideband regions require $|\widehat M - M_{\Ds}| < 12 \textnormal{ MeV}/c^2$.  The signal region is $|\Delta M| < 30\textnormal{ MeV}/c^2$, while the sideband region is $50 < |\Delta M| < 140\textnormal{ MeV}/c^2$.  Efficiencies for DT modes range from 0.3\% to 38\%.

The \Ds\ decay final states under consideration can often be reached through multiple intermediate resonances.  For example, in our Monte Carlo the final state $\Km\Kp\pip$ is an incoherent mixture of $\Kstarzb \Kp$ (43\%), $\phi \pip$ (38\%), $\overline{K}_0^*(1430)^0 \Kp$ (8\%), nonresonant production (7\%) and $f_0(980) \pip$ (4\%).  The reconstruction efficiency can depend significantly on which resonances are produced.  Knowledge of the relative contributions of these intermediate states is incomplete.  We compare invariant mass distributions of pairs of \Ds\ daughters in data and Monte Carlo, and use the resulting information on resonant structures to reweight the assumed intermediate state components.  The resulting excursions in the efficiency are taken as systematic uncertainties.  Where there is a significant component that cannot be explicitly assigned to any intermediate state, we find the worst-case variations between the dominant components.  As an illustration, for \Km\Kp\pip\, we find that $\phi\pip$ and $\Kstarzb \Kp$ have very similar (and lowest) efficiencies, while the nonresonant component is 7\% higher and the others lie between these extremes.  By selecting on the $\Km\Kp$ and $\Km\pip$ invariant masses we ascribe 90\% of reconstructed events to $\phi\pip$ or $\Kstarzb \Kp$; varying the assumed efficiency for the remaining events within the limits above changes the inferred average efficiency, leading to a systematic uncertainty of 1.5\%.  The uncertainties assigned vary from zero for the two-body final states to $6\%$ for \Km\Kp\pip\piz\ (where there is a large efficiency difference between $\phi\rho^+$ and $\overline{K}^{*0} K^{*+}$).  We also include uncertainties in the PDG 2007 fit values for $\Br(\eta \to \gamma \gamma)$ (0.7\%) and $\Br(\eta' \to \pip\pim\eta)$ (3.1\%), and correct for the difference between the PDG fit for $\Br(\KS \to \pip\pim)$ and the value used in \textsc{geant}.

Systematic uncertainties for the simulation of track, \KS, $\piz$, and $\eta$ reconstruction and PID efficiencies are determined using partial versus full reconstruction of events in \cleoc's $\psi(2S)$ and $\psi(3770)$ datasets; the methods are shared with the \Dz/\Dp\ branching fraction analysis \cite{Dobbs2007}.  Tracking efficiencies are verified using $\psi(3770) \to \D\Dbar$ events for $\pi^\pm$ and $K^\pm$, and using $\psi(2S) \to \pip\pim\Jpsi$ for $\pi^\pm$.  Good agreement is found, and an uncertainty of 0.3\% per track is used,
correlated among all tracks, with an additional uncertainty of 0.6\% per kaon
added in quadrature.  
Systematic effects in the PID efficiency are studied using $\psi(3770) \to \D\Dbar$ events; in general data has slightly lower efficiency than the simulations and corrections are applied.  Because the corrections are momentum-dependent this is also affected by the uncertainty on the intermediate resonant states.  The corrections applied range from $(-0.2 \pm 0.2)\%$ for $\pip\eta$ to $(-3.7\pm 1.4)\%$ for \Km\Kp\pip\piz.
Neutral kaon efficiencies are verified using \D\Dbar\ events and the $\Dsp\to\KS\Kp$ mode; a systematic uncertainty of 1.9\% per \KS\
candidate is used.  The $\piz$ efficiency is checked with $\psi(2S)\to\piz\piz\Jpsi$ decays, and the $\eta$ efficiency with $\psi(2S)\to\eta\Jpsi$ events.  In both cases there are discrepancies between data and the simulation, and relative corrections of $(-3.9 \pm 2.0)\%$ per \piz\ and $(-5.7 \pm 4.0)\%$ per $\eta$ are applied.  

The nominal signal lineshapes used in the ST yield fits are derived from the simulation, and the backgrounds are either linear or quadratic.  We determine systematic uncertainties in the yields by relaxing each assumption separately: the mass resolution is allowed to vary by an overall scale factor, and the background is parameterized by a second-order polynomial if the nominal fit uses a linear one, or vice versa.  The size of the resulting excursions vary from 0.2\% (\Km\Kp\pip) to 8.6\% (\Km\Kp\pip\piz) for background shape and 0.1\% (\KS\Kp) to 10.3\% ($\pip\eta$) for width.

The efficiency for a reconstructed DT event to lie in the signal region depends on the mass resolutions for both candidates.  Errors in modeling the resolution will thus cause errors in the DT efficiency which are correlated with the ST signal lineshape uncertainties.  To estimate this effect we use the best fit results from the ST width check to determine the changes expected in the DT efficiency. The difference due to each decay mode is taken as a systematic uncertainty competely correlated with the corresponding ST uncertainty.  The range of these effects is 0--8\%.

In addition, we consider mode-dependent systematic uncertainties arising from our modeling of average $\Ds^*\Ds$ event multiplicity and detector noise (0--3\%), the final state radiation spectrum generated by \textsc{photos} (0.2--1.2\%), and our simulation of initial state radiation (0--0.8\%).

\begin{table*}
\caption{\label{tbl:bfresults}Branching fraction results from this
  analysis, world average branching fractions from the PDG 2007 fit \cite{Yao:2006px}, ratios of branching fractions to $\Br(\Dsp\to\Km\Kp\pip)$, and charge asymmetries $\mathcal{A}_{CP}$.  Uncertainties on CLEO-c measurements are statistical and systematic, respectively.}
\begin{center}
\begin{tabular}{lcccc}
\hline\hline
Mode & This Result \Br\ (\%) & PDG 2007 fit \Br\ (\%) & $\Br/\Br(\Km\Kp\pip)$ & $\mathcal{A}_{CP}$ (\%)\\
\hline
\KS\Kp & \BFKSK & $2.2 \pm 0.4$ & \BRKSK & \AKSK \\
\Km\Kp\pip & \BFKKpi & $5.3 \pm 0.8$ & 1 & \AKKpi\\
\Km\Kp\pip\piz & \BFKKpipiz & --- & \BRKKpipiz & \AKKpipiz\\
\KS\Km\pip\pip & \BFKSKpipi & $2.7 \pm 0.7$ & \BRKSKpipi & \AKSKpipi\\
\pip\pip\pim & \BFpipipi & $1.24 \pm 0.20$ & \BRpipipi & \Apipipi\\
$\pip\eta$ & \BFpieta & $2.16 \pm 0.30$ & \BRpieta & \Apieta\\
$\pip\eta'$ & \BFpietaprime & $4.8 \pm 0.6$ & \BRpietaprime & \Apietaprime\\
\Kp\pip\pim & \BFKpipi & $0.67 \pm 0.13$ & \BRKpipi & \AKpipi\\
\hline\hline
\end{tabular}
\end{center}
\end{table*}

Peaking backgrounds in ST events are found to be negligible compared to the size of the background shape uncertainties.  Very small crossfeeds (of order 0.5\% or less) are expected between various DT modes and are included in the fit; peaking DT backgrounds from other sources mostly arise from $D^*D^*$ reflections and are again found to be negligible.

Systematic uncertainties are propagated to the final results by altering fit inputs (efficiencies and yields) with appropriate correlations and noting the variations in the results.  The analysis was validated on a simulated generic sample of open charm production with thirty times the statistics of the data, and successfully reproduced the input branching fractions.

We have separate yields and efficiencies for \Dsp\ and \Dsm\ events, so it is possible to compute asymmetries
\[ \mathcal{A}_{CP,i} = \frac{y_i/\epsilon_i - y_{\bar \imath}/\epsilon_{\bar \imath}}
 {y_i/\epsilon_i + y_{\bar \imath}/\epsilon_{\bar \imath}},
\]
which are sensitive to direct $CP$ violation in \Ds\ decays (expected to be very small in the Standard Model).  Most systematic uncertainties cancel in this ratio; the ones that remain are due to charge dependence in tracking and PID, and the dependence of the ST yields on the signal lineshape and background parametrization.

The obtained branching fractions, branching ratios, and $CP$ asymmetries are shown in Table~\ref{tbl:bfresults}. The values we obtain are consistent with the world averages \cite{Yao:2006px} and significantly more precise than any previous absolute measurements of \Ds\ branching fractions.  This is also the first result where all eight modes are measured simultaneously; the PDG fit combines many disparate branching ratio results.  No significant $CP$ asymmetries are observed.  We additionally obtain the number of $\Ds^*\Ds$ events $N_{\Ds^*\Ds} = \NDsDs$, which gives $\sigma_{\Ds^*\Ds}(4.17\textnormal{ GeV}) = (\sigDsDs)$~nb; in order, the uncertainties are statistical, systematic due to this measurement, and for the cross-section, systematic due to luminosity measurement \cite{Dobbs2007}.  The cross-section is consistent with earlier \cleoc\  results obtained via a scan of this energy region \cite{CroninHennessy:2008yi}.

\begin{table}
\caption{\label{tbl:partialbf} Partial branching fractions $\Br_{\Delta M}$ for $\Km\Kp\pip$ events with $\Km\Kp$ mass within $\Delta M$ MeV$/c^2$ of the $\phi$ mass. Uncertainties are statistical and systematic, respectively.}
\begin{center}
\begin{tabular}{lc}
\hline\hline
Value & This Result \Br\ (\%) \\
\hline
$\Br_{5}$ & \BFphipiv\\
$\Br_{10}$ & \BFphipix\\
$\Br_{15}$ & \BFphipixv \\
$\Br_{20}$ & \BFphipixx\\
\hline\hline
\end{tabular}
\end{center}
\end{table}

\begin{figure}
 \includegraphics*[width=\linewidth]{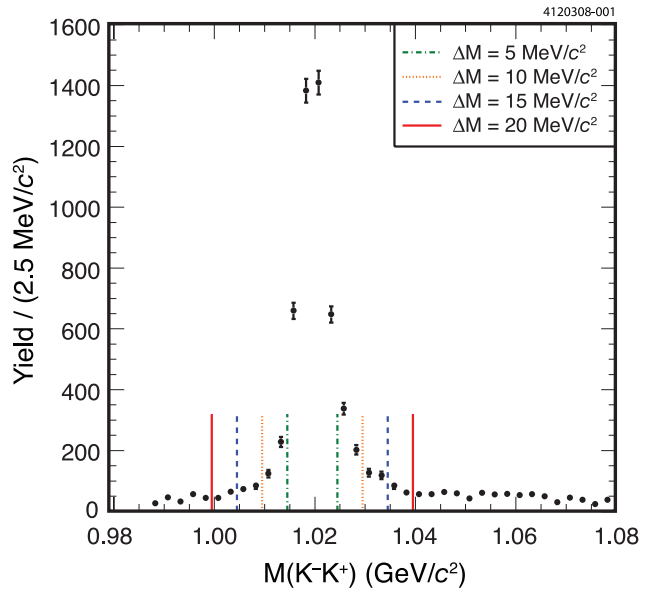}
\caption{\label{fig:kkmassfit}(Color online) Yields of $\Dspm \to K^\mp K^\pm \pi^\pm$ single tag events
versus $\Km\Kp$ invariant mass; no efficiency corrections have been applied.  The ST fit procedure for the full $\Km\Kp\pip$ sample is applied here to the subsample of each bin of $M(\Km\Kp)$ and the resulting yields plotted, hence backgrounds have been subtracted and the yields shown are signal.  A $\phi$ peak is visible above an additional broad signal component.  The lines show the mass window boundaries for the partial branching fractions in Table~\ref{tbl:partialbf}. }
\end{figure}

A quantity conventionally termed $\Br(\Dsp\to\phi\pip)$ has often been used as a reference branching fraction for \Dsp\ decays; operationally it is measured by making kinematic selections on the kaon pair in $\Dsp \to \Km\Kp\pip$ events and assuming a pure $\phi \to \Km\Kp$ signal.  However, the Dalitz plot for this mode shows the presence of a significant broad scalar component under the $\phi$ peak, whose contribution to the observed yield varies from less than 5\% to over 10\% depending on the $\phi$ candidate selection criteria.  Figure~\ref{fig:kkmassfit} shows the mass spectrum of $\Dsp\to\Km\Kp\pip$ events in this mass region; when fit by a single Gaussian, the $M(\Km\Kp)$ resolution is $1.1\textnormal{ MeV}/c^2$.  The scalar component will additionally interfere with the $\phi$ contribution, altering the observed rate of events in the $\phi$ peak from the $\Dsp \to \phi\pip$ fit fraction which would be measured in an amplitude analysis.  These variations are comparable to or exceed the systematic uncertainties in our measurements.  For this reason, we do not quote a branching fraction for the resonant mode $\Dsp \to \phi\pip$; this quantity can only be unambiguously measured with an amplitude analysis, which is of limited utility as a reference branching fraction.  We instead provide partial branching fractions $\Br_{\Delta M}$, which are defined as the branching fraction for $\Km\Kp\pip$ events where the $\Km\Kp$ pair satisfies $|M(\Km\Kp) - 1019.5\textnormal{ MeV}/c^2| < \Delta M$~(MeV$/c^2$); the values obtained are listed in Table~\ref{tbl:partialbf}.  The systematic uncertainties quoted for $\Br_{\Delta M}$ include uncertainties due to resolution.  We emphasize that these are not measurements of the quantity $\Br(\Dsp \to \phi\pip \to \Km\Kp\pip)$, but are intended as references to normalize other \Dsp\ branching fractions when most of the $\Km\Kp\pip$ phase space must be excluded for background reasons.

In summary, we have presented the first application of a tagging technique at a center of mass energy of 4.17 GeV to measure eight hadronic \Dsp\ branching fractions with precisions exceeding world averages.  For the key mode $\Dsp \to \Km\Kp\pip$, the statistical and systematic uncertainties are comparable.  As the experimental meaning of $\Br(\Dsp\to\phi\pip)$ is ill-defined at this level of precision without a full amplitude analysis, we do not report it.  We instead provide partial branching fractions for windows centered on the $\phi$ mass which do not assume a specific resonant composition of the decay. 

We gratefully acknowledge the effort of the CESR staff
in providing us with excellent luminosity and running conditions.
This work was supported by
the A.P.~Sloan Foundation,
the National Science Foundation,
the U.S. Department of Energy,
the Natural Sciences and Engineering Research Council of Canada, and
the U.K. Science and Technology Facilities Council.

\end{document}